\documentclass[aps,prd,preprint,floatfix,nobibnotes,nofootinbib]{revtex4-2}
\usepackage{amsmath}
\usepackage{amssymb}
\usepackage{graphicx}
\usepackage{float}
\usepackage{ulem}
\usepackage[colorlinks=true]{hyperref}
\usepackage{slashed}
\usepackage{pgffor}

\newcommand{\be}{\begin{equation}}
\newcommand{\ee}{\end{equation}}
\newcommand{\ba}{\begin{eqnarray}} 
\newcommand{\ea}{\end{eqnarray}} 
\newcommand{\nn}{\nonumber}

\newcommand{\bea}{\begin{eqnarray}}
\newcommand{\eea}{\end{eqnarray}}

\usepackage[usenames, dvipsnames]{xcolor}
\definecolor{darkgreen2}{RGB}{0, 150, 0}

\usepackage{caption}
\usepackage{subcaption}
\usepackage{float}
\numberwithin{equation}{section}

\usepackage{setspace}

\begin{document} 

\title{The gauge invariance of non-perturbative vertex prescriptions}

\author{M.E. Carrington}
\email[]{carrington@brandonu.ca} 
\affiliation{Department of Physics, Brandon University, Brandon, Manitoba, R7A 6A9 Canada}
\affiliation{Department of Physics \& Astronomy, University of Manitoba, Winnipeg, Manitoba, R3T 2N2 Canada}
\affiliation{Winnipeg Institute for Theoretical Physics, Winnipeg, Manitoba}
 
\author{A.R. Frey}
\email[]{a.frey@uwinnipeg.ca} \affiliation{Department of Physics, University of Winnipeg, Winnipeg, Manitoba, R3M 2E9 Canada}
\affiliation{Department of Physics \& Astronomy, University of Manitoba, Winnipeg, Manitoba, R3T 2N2 Canada}
\affiliation{Winnipeg Institute for Theoretical Physics, Winnipeg, Manitoba}

\author{B.A. Meggison}
\email[]{brett.meggison@gmail.com} 
\affiliation{Department of Physics \& Astronomy, University of Manitoba, Winnipeg, Manitoba, R3T 2N2 Canada}
\affiliation{Winnipeg Institute for Theoretical Physics, Winnipeg, Manitoba}

\begin{abstract}
We study the gauge invariance of different continuum methods to include non-perturbative effects in gauge theories. We work with three dimensional quantum electrodynamics and implement vertices using two different methods: a set of coupled Schwinger-Dyson (SD) integral equations, and the self-consistent equations obtained from the three loop 3-particle irreducible (3PI) effective action. We work in Landau gauge and assess the extent to which results are gauge invariant by checking how well the Ward identity is satisfied. Our results show that there is a fairly significant violation of the Ward identity at large coupling, although the 3PI effective theory is slightly better than 
the SD vertex. We also compare the results of both calculations with the commonly used Ball-Chiu ansatz and show that the agreement of the ansatz with both non-perturbative vertices is fairly good at small coupling but deviates more significantly at large coupling. We compare results for the two point functions of the theory and discuss the possible implications for phase transitions. 
\end{abstract}

\maketitle
\newpage

\section{Introduction}
\label{sec-introduction}

The purpose of this paper is to study the gauge invariance of different non-perturbative methods that are commonly used to calculate vertex functions in strongly coupled field theories. Vertex functions are notoriously difficult to calculate and the problem is commonly avoided by introducing some simplified ansatz to represent the vertex. The most popular choice is the Ball-Chiu (BC) ansatz \cite{BallChiu1} which satisfies the Ward identity by construction, but does not necessarily correctly represent the transverse vertex components. 

The primary technical difficulty in the calculation of non-perturbative vertices is that the number of independent variables in the phase space of an $n$-point function increases with $n$. In a covariant theory (at zero temperature and in the absence of anisotropy) a 2-point function depends on one variable in momentum space. For example, a two point function $\tilde G(x,y)$ takes the momentum space form $G(p)$ and depends only on the magnitude of $|p|$. In comparison a momentum space three point function depends on three variables. For example, a three point function $\tilde\Gamma(x,y,z)$ has the momentum space form $\Gamma(p,k)$ and depends on $|p|$ and $|k|$ and the angle between the two vectors. The larger phase space (three independent variables instead of one) dramatically increases the requirements for both memory and compute time in the calculation of vertex functions.

In addition, gauge theories are much more difficult to work with than scalar theories. For a theory that describes particles with spin and/or fermion number, each $n$-point function has spinor/tensor structure. In practice one introduces an appropriate basis of bi-spinors and decomposes the $n$-point function into scalar coefficients. A familiar example is the electron self-energy, which can be written in terms of two scalar functions corresponding to wavefunction and mass renormalization (see eq.~(\ref{sigma-eq})). In comparison, the vertex function in quantum electrodynamics (QED) has 12 independent components, each of which is determined from its own integral equation. The convergence of a large set of coupled equations is well known to be a numerically challenging task. 

In this paper we will calculate the non-perturbative QED vertex in three Euclidean dimensions using two different methods. The most common and in most ways the simplest method to calculate non-perturbative vertex functions is to solve the set of Schwinger-Dyson (SD) equations obtained from the action (see the review \cite{Roberts:1994dr}). The SD equations form an infinite coupled hierarchy and therefore must be truncated. The simplest method is to choose a value $n_{\rm max}$ and set the $n_{\rm max}$-point function to its perturbative form, which leaves a closed set of coupled SD equations for the lower $n$-point functions. It is clear that this kind of truncation, for any value of $n_{\rm max}$, will violate gauge invariance. More complicated methods have been proposed to truncate the SD equations without violating gauge invariance (see \cite{Binosi:2009qm}), but  calculations using these more sophisticated truncations are much more difficult. Part of the goal of this work is to evaluate the extent to which a naive truncation of the SD equations violates gauge invariance. In this work we use a bare four point function (which is zero in QED) and solve a set of coupled equations for the two- and three point functions of the theory. In addition we will calculate the non-perturbative vertex obtained from the three loop three-particle irreducible (3PI) effective action.  This vertex also violates gauge symmetry identities \cite{Calzetta:2004sh,Reinosa:2009tc}, but it can be proven that gauge invariance is respected to the level of the truncation \cite{Arrizabalaga:2002hn,Carrington:2003ut}. We comment that strategies to push the gauge violation in $n$PI theories to even higher orders have been studied \cite{Brown:2016vaj}. For a review of non-perturbative functional methods in QCD see \cite{Huber:2018ned}.

For simplicity we work only in Landau gauge. In spite of the fact that the gauge is fixed, we can evaluate the gauge invariance of the prescription by checking how closely the non-perturbative vertex satisfies the Ward identity. Our results show that at the largest coupling we used ($\alpha=5$) the Ward identity is violated by approximately 30\%.  We emphasize that the effect of this gauge dependence on a physical quantity, like the fermion condensate, can only be tested by performing calculations in a different gauge. A direct check by comparison of the Landau and Feynman gauge results is in progress and will be presented in a future publication. 

We will also compare both the SD and 3PI vertices with the BC vertex ansatz. 
The idea underlying the BC ansatz is to find a representation of the three point functions in a gauge theory in terms of two point functions. This ansatz is constructed to satisfy the corresponding Ward identity, so in this sense there is no issue with gauge invariance. 
The problem is that the Ward identity constrains only the longitudinal components of the vertex, which means that  there is no way to evaluate the importance or accuracy of the transverse components. 
Another ansatz we have not considered is the Curtis-Pennington vertex \cite{pennington0}. This ansatz was constructed to guarantee multiplicative renormalisability in general linear covariant gauges, but in our calculation (using Landau gauge in three dimensions) this is not an issue. 
Furthermore, the Curtis-Pennington construction was originally performed in quenched QED
and it was later shown by Pennington and Kizilersu \cite{Curtis:1993py,Kizilersu:2009kg}
that the transverse Curtis-Pennington term is actually not a good choice for the vertex in the
fermion loop of the photon SD equation. 

The motivation for using a vertex ansatz is that, as explained above, the number of independent variables in a two point function is smaller and this reduces the numerical difficulty by a huge factor. For this reason many non-perturbative calculations are done using some type of vertex ansatz (some interesting examples include \cite{Goecke:2008zh,Kizilersu:2014ela,Eichmann:2021zuv,Albino:2022efn}), but there is no way of confirming whether or not the dominant physics is correctly captured by the ansatz. 
By comparing the SD and 3PI vertices with the BC ansatz we are able to assess the accuracy of these commonly used approximations. 

This paper is organized as follows. In sec.~\ref{notation-sec} we define our notation and discuss the structure of the integral equations that we will solve. In sec.~\ref{sec-method} we describe our numerical method.  In sec.~\ref{sec-results} we present our results and discuss their importance. In sec.~\ref{sec-conclusions} we discuss possible future directions of this research and present our conclusions. Some details are given in appendix \ref{ap-1loop}. 

\section{Notation}
\label{notation-sec}
We use natural units ($\hbar=c=1$). 
We work in three dimensional Euclidean space. 
Greek letters denote Lorentz indices of three-dimensional vectors.
We use a representation of three four-dimensional gamma matrices that satisfy the commutation relation $\{\gamma^\mu,\gamma^\nu\}=2\delta_{\mu\nu}$. A three point vertex function depends on two external momenta because of momentum conservation. We write $\Gamma(k,p)$ where $p$ and $k$ are the three momenta of the incoming and outgoing fermions, respectively.  The function depends on the magnitude of the two momenta and the angle between them, which we call $\theta$, and this is not shown explicitly in our notation. 
We define the combinations
\bea
q^\mu &=& k^\mu-p^\nu \nn \\
l^\mu&=& k^\mu+p^\mu \nn \\
\triangle &=& p^2 k^2 - \left(p \cdot k\right)^2 
\eea
which will be used to simplify some equations. In some cases to save space we represent functional arguments with subscripts, for example $A(p)\equiv A_p$.

One advantage of working in three dimensions is that renormalization is trivial. The only divergence is a linear divergence in the photon polarization tensor that can be removed with a subtraction. We define 
\bea
\tilde{\Pi}(p) &=& \Pi(p) - \Pi(0) 
\eea
and use the renormalized $\tilde{\Pi}(p)$ in all numerical calculations. 

\subsection{Two point functions}
\label{2point-sec}

The inverse fermion propagator is written
\bea
S^{-1}(p) &=&  S^{-1}_0(p) + \Sigma(p) \, ,
\label{Sinv-def}
\eea
where $\Sigma(p)$ is the self-energy and the bare fermion propagator is
\bea
S_0^{-1}(p) &=&   -i \slashed{p} + m\,.
\eea
The inverse photon propagator has the form
\bea
G_{\mu \nu}^{-1}(p) &=& G_{0\,\mu \nu}^{-1}(p) + \Pi_{\mu \nu} (p) \, ,\label{Geqn0}
\eea
where $\Pi_{\mu \nu} (p)$ is the polarization tensor 
and the bare inverse propagator is 
\bea
G_{0\,\mu \nu}^{-1}(p) &=& p^2 \left( P_{\mu \nu}^T(p)+\frac{1}{\xi}P_{\mu \nu}^L(p) \right) \label{Geqn} 
\eea
with the transverse and longitudinal tensors defined as 
\bea
P_{\mu \nu}^T(p) &=& \delta_{\mu \nu}-\frac{p_\mu p_\nu}{p^2} \label{tran}\\
P_{\mu \nu}^L(p) &=& \frac{p_\mu p_\nu}{p^2} \,.\label{long}
\eea
We work in Landau gauge $\xi=0$.

\begin{figure}
	\begin{centering}
\begin{subfigure}{0.45\textwidth}
	\centering
		\includegraphics{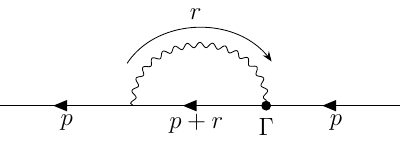}
		\subcaption{fermion self-energy diagram}
		\label{fermionSelfEnergy}
\end{subfigure}
\begin{subfigure}{0.45\textwidth}
	\centering
	\includegraphics{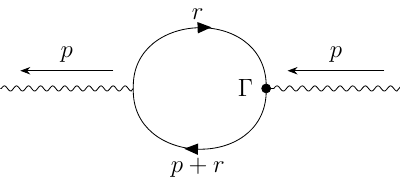}
	\subcaption{photon self-energy diagram}
	\label{photonSelfEnergy}
\end{subfigure}
\caption{propagator corrections}
\label{selfenergies}
\end{centering}
\end{figure}

Using either a SD approach or the three loop 3PI effective action, the fermion self-energy and photon polarization tensor are 
\bea
\Sigma(p) &=& e^2 \int \frac{d^3 r}{(2 \pi)^3} \gamma_\mu S(p+r) \Gamma_\nu (p+r,p) G_{\mu \nu}(r) \label{sigma}\\
\Pi_{\mu \nu}(p) &=& - e^2 \int \frac{d^3 r}{(2 \pi)^3} {\rm Tr}\left[ \gamma_\mu S(p+r) \Gamma_\nu(p+r,r) S(r) \right] \,. \label{pi}
\eea
These equations are represented in figures \ref{fermionSelfEnergy} and \ref{photonSelfEnergy}. 
We emphasize that although these equations have exactly the same form when using either SD or 3PI equations, it is not true that the two point functions themselves are the same, because in each case the  vertex $\Gamma$ is defined differently. This is explained in detail in the next section. 
We note that the equations of motion for the 3PI two point functions have the same form as the corresponding SD equations because of the structure of the equation for the 3PI vertex function (see \cite{Carrington:2004sn,Berges:2004pu}). 

The fermion self-energy is a bi-spinor and the photon polarization tensor is a $3\times 3$ tensor because it carries Lorentz indices. These quantities can be written in terms of scalar functions by performing a tensor decomposition.  
The fermion self-energy is written in terms of two scalar functions as
\bea
\Sigma(p) &=& -i \slashed{p} A(p)  + B(p)\, ,
\label{sigma-eq}
\eea 
which means that the non-perturbative fermion propagator has the form
\bea
S^{-1}(p) &=&   -i (1+A(p))\slashed{p} + m+B(p) \,.
\label{S-nonpert}
\eea
The polarization tensor is transverse and therefore can be written in terms of one scalar component as
\bea
\Pi_{\mu \nu} (p) &=& P_{\mu \nu}^T(p) \Pi(p) \label{Pieqn}.
\eea

\subsection{Three point functions}
\label{threepointfunctions-subsec}

For a three point function the structure of the spinor/tensor decomposition is much more complicated than it is for a two point function \cite{BallChiu1,pennington-2,Eichmann:2026ttr}. For the QED three point function there are 12 independent structures. This is easy to understand by noting that there are 12 independent functions of the two external momenta that have the correct Dirac and Lorentz structure, which can be represented as
\bea
\{\gamma_\mu , p_\mu, k_\mu  \} \times \{ 1,\slashed{p},\slashed{k},\slashed{p}\slashed{k} \}.
\label{naif}
\eea
In practice one does not work directly with these bi-spinor/tensor combinations but rather defines 12 independent linear combinations that are easier to work with. 
In general one wants to choose basis functions that are (as far as is possible) orthogonal to each other so that the integral equations that determine the scalar coefficients are as
much as possible decoupled from each other. 
In addition, it is crucial to choose basis tensors so that the vertex dressing functions are free from kinematic singularities. 
We comment that for the two point functions discussed in sec.~\ref{2point-sec} the choice of basis tensors was essentially trivial, and the situation is more complicated for vertex functions because of the two independent momenta. 
The simplest approach is to work with combinations of the tensors in (\ref{naif}) that are transverse and longitudinal to the relative momentum $q_\mu = k_\mu - p_\mu$. Several different definitions have been used in the literature (see, for example, \cite{Williams:2015cvx,davy}). 
We will use the eight transverse tensors defined by Ball and Chiu \cite{BallChiu1}, and four additional longitudinal combinations. Our definitions are
\bea
T_1^\mu(k,p) &=& \left(p^\mu k \cdot q - k^\mu p \cdot q\right) 
=2p_\nu k_\lambda \left(\delta^{\mu[\nu}q^{\lambda]}  \right)\nn \\
T_2^\mu(k,p) &=& T_1^\mu(k,p) (\slashed{p}+\slashed{k}) \nn \\
T_3^\mu(k,p) &=& q \cdot q \gamma^\mu - q^\mu \slashed{q} 
=2q_\nu q_\lambda\left( \gamma^{[\mu}\delta^{\nu]\lambda}\right)\nn \\
T_4^\mu(k,p) &=& -\frac{1}{2} T^\mu_1(k,p) (\slashed{p}\slashed{k}- \slashed{k}\slashed{p}) 
=T_1^\mu(k,p)p_\nu k_\lambda \left(\gamma^{[\nu}\gamma^{\lambda]}\right)\nn \\
T_5^\mu(k,p) &=& \frac{1}{2} (\gamma^\mu \slashed{q} -\slashed{q}\gamma^\mu) 
=q_\nu\left(\gamma^{[\mu}\gamma^{\nu]}\right)\nn \\
T_6^\mu(k,p) &=& \gamma^\mu (k^2-p^2)-(p+k)^\mu\slashed{q} 
=2q_\nu l_\lambda\left( \gamma^{[\mu}\delta^{\nu]\lambda}\right)\nn \\
T_7^\mu(k,p) &=& \frac{1}{2}(k^2-p^2)(\gamma^\mu \slashed{p}+\gamma^\mu \slashed{k}-p^\mu - k^\mu)+\frac{1}{2}(k^\mu+p^\mu)(\slashed{p}\slashed{k}-\slashed{k}\slashed{p}) =-\frac 14 l_\nu l_\lambda q_\rho \delta^{\nu[\mu}\left[\gamma^{\rho]},\gamma^\lambda\right]\nn \\
T_8^\mu(k,p) &=& p^\mu \slashed{k}-k^\mu \slashed{p} + \frac{1}{2}(\gamma^\mu \slashed{k}\slashed{p}-\gamma^\mu \slashed{p}\slashed{k}) 
=-p_\nu k_\lambda \gamma^{[\mu}\gamma^\nu\gamma^{\lambda]}\label{T8O}\nn \\
L_1^\mu(k,p) &=& \slashed{q} q^\mu \label{L1} \nn \\
L_2^\mu(k,p) &=& (k^2-p^2) q^\mu \nn \\
L_3^\mu(k,p) &=& (k^2-p^2) (\slashed{k}+\slashed{p}) q^\mu  \label{L3}\nn \\
L_4^\mu(k,p) &=& (\slashed{k}\slashed{p}-\slashed{p}\slashed{k}) q^\mu \label{our-basis}\,.
\eea
It is easy to check that the bi-spinors/tensors $T_i^\mu$ are transverse with respect to the vector $q_\mu$  and the $L_i^\mu$ bi-spinors/tensors are longitudinal: $T^\mu_i(k,p) P^L_{\mu \nu}(q) = 0$ and $L^\mu_i(k,p) P^T_{\mu \nu}(q)=0$ for all $i$, where $P^L_{\mu \nu}$ and $P^T_{\mu \nu}$ are defined in eq.~(\ref{long}). 
When given, the second expression for each spinor/tensor structure makes the gamma matrix structure explicit and emphasizes the relationship between the different vertex structures. Square brackets around indices are weighted antisymmetrization, so $a^{[\mu\nu]}\equiv (a^{\mu\nu}-a^{\nu\mu})/2$ and $b^{[\mu\nu\lambda]} = (b^{\mu[\nu\lambda]}+b^{\nu[\lambda\mu]}+b^{\lambda[\mu\nu]})/3$ for any tensors $a^{\mu\nu}$ and $b^{\mu\nu\lambda}$. 
We note that this basis for bi-spinor/tensor structures includes known multipole operators such as the dipole moment $T_5^\mu$ and the anapole (toroidal multipole) moment $T_3^\mu$ (see \cite{anapole}).

The vertex dressing functions are defined by writing the non-perturbative vertex as a sum of scalar functions times each of these basis vertex structures: 
\bea
\Gamma^\mu(k,p) &=& \gamma^\mu +\delta \Gamma^\mu(k,p) \nn \\
\delta\Gamma^\mu(k,p) &=&  \sum_{i=1}^8 f^\tau_i \tau_i(k,p) T_i^\mu + \sum_{i=1}^{4} f^\lambda_i \lambda_i(k,p) L_i^\mu \label{Gam} 
\eea
where we include the coefficients
\bea
f^\tau_i &=& \{ i,1,1,i,i,1,i,1 \} \nn\\
f^\lambda_i &=& \{ 1,i,1,i \}
\eea
so that all dressing functions are real. As discussed above, all vertex dressing functions depend on three variables: the magnitudes of the two momenta $p$ and $k$ and the angle between them. 

To calculate the vertex dressing functions, we need to have an expression for the non-perturbative vertex. We will study the 3PI and SD definitions. The 3PI vertex is given in eq.~(\ref{Gam}) with \cite{Berges:2004pu,Carrington:2007ea} 
\bea
\delta \Gamma^\mu(k,p) &=& -e^2 \int \frac{d3d r}{(2 \pi)^3}\Gamma^\nu(k,k+r)S(k+r)\Gamma^\mu(k+r,p+r)  S(p+r)\Gamma^\rho(p+r,p)G^{\nu \rho}(r) \nn\,.\\\label{delGam}
\eea
The SD approach gives two different integral equations for the vertex correction $\delta\Gamma$ which come from the two different master equations. One is the vertex in eq.~(\ref{Gam}, \ref{delGam}) with the third vertex in (\ref{delGam}) set to the bare vertex ($\Gamma^\rho(p+r,p)\to\gamma^\rho$) and the other is with the second vertex set to the bare vertex ($\Gamma^\mu(k+r,p+r)\to\gamma^\mu$). These two expressions make two different non-perturbative vertices that are used respectively in the integral equations (\ref{sigma}) and (\ref{pi}). 

\begin{figure}
	\centering
	\includegraphics{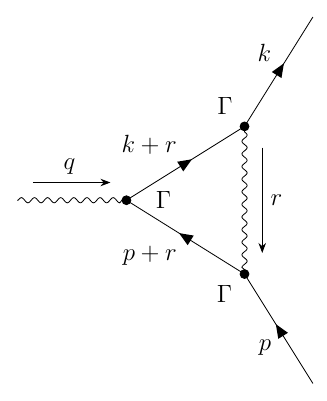}
	\caption{The vertex correction $\delta\Gamma_\mu$ in eq.~(\ref{delGam}). }
	\label{vertexCorrection}
\end{figure}

\subsection{Scalar equations}
\label{scalar-sec}

The next task is to calculate scalar integral equations for the dressing functions. To do this we calculate the projections of the equations that define them
\bea
&& {\rm Tr}\left[\slashed{p}((\ref{sigma})-(\ref{sigma-eq}))\right] =0\nn \\
&& {\rm Tr}\left[(\ref{sigma}) - (\ref{sigma-eq})\right] =0 \nn \\
&& P^T_{\mu\nu}\left[(\ref{pi})-(\ref{Pieqn})\right] =0 \nn \\
&& {\rm Tr}[T_i^\mu ((\ref{Gam})-(\ref{delGam}))] =0 \text{~for~}i\in{1,8} \nn \\
&& {\rm Tr}[L_i^\mu ((\ref{Gam})-(\ref{delGam}))] =0 \text{~for~}i\in{1,4}
\eea
where the traces are over Dirac indices and the numbers indicate the right sides of the corresponding equations. Note that in the last three equations, the Lorentz indices are contracted with the free indices in the indicated equations. This set of 15 independent equations can be solved for the 15 dressing functions. The result is a coupled set of self-consistent equations that can be  solved numerically by initializing at some guessed form and iterating until a solution is found. 

Although the process is straightforward the calculation itself is very difficult. %
We generated the scalar equations (performed all traces and contracted over Lorentz indices) and found the integral equations for each dressing function using FORM \cite{form}. The resulting integral equation for a single vertex dressing function produces a text file that is on average approximately $30$~MB.
In addition these equations have many terms with factors of the form $(p-k)^2=q^2$ and $\triangle = p^2 k^2-(p k \cos(\theta))^2$ (and powers of these expressions) in their denominators. The presence of these factors appears to indicate singularities in regions where $p\to k$ and $\theta\to (0,\pi)$. 
In fact there are no singularities. This is the motivation behind the particular structure of the Ball-Chiu basis: it is designed so that all vertex dressing functions are divergence free. The result can be proven by analysing the structure of the vertex integral equations at the one loop level. One can substitute bare lines and vertices and expand around the suspected singular points and verify that all functions are in fact finite \cite{BallChiu1}. While there are other bases that also give finite dressing functions, the important point is that an arbitrary selection of the 12 independent tensors in (\ref{naif}) will not (the basis conversion can contain coefficients that diverge at some points in parameter space).

We make two other comments. In this work we have set $\lambda_4$ to zero. The reason is that we know that $\lambda_4$ must be zero if the Ward identity that relates the three vertex to the fermion self-energy is satisfied (see eq.~(\ref{wardIdentity})). This can be easily seen from the fact that $q^\mu L_4^\mu$ is proportional to a commutator of gamma matrices, and this structure is not present in the fermion self-energy. 
We also comment that the authors of \cite{pennington-2,davy} found that redefining $T_4^\mu$ as $\tilde{T}_4^\mu$ as
\be
\tilde{T}_4^\mu = \frac{2}{k^2-p^2}\left( 2 T^\mu_{4} + q^2 T^\mu_{7}\right) \label{tildeT4Davy}
\ee
reduced singular behaviour, but we found no advantage to this definition. 

As explained above, we know from their one loop structure that the vertex dressing functions we have defined are finite. It is clear however that problems will arise in a numerical calculation. When finite results should be obtained by summing many terms that are individually very large, results can be very inaccurate. When solving a large set of self-consistent integral equations this problem will manifest as instability in the convergence procedure. In practical terms, when one attempts to solve the equations that are obtained from the FORM procedure described above, it is impossible to find a numerical solution. 
In section \ref{sec-method} we will give more details on our numerical method. For now we emphasize that the success of the procedure depends critically on implementing an efficient method to simplify the set of equations obtained from the initial FORM output so that singular factors in denominators are cancelled as completely as possible before numerical calculations are started. 

\subsection{Gauge invariance}

As discussed in section \ref{sec-introduction}, the three loop 3PI effective theory and the set of SD integral equations truncated by setting the bare four vertex to the perturbative form are not gauge invariant. It is commonly argued that the gauge dependence introduced with these methods is small. The extent to which this is actually true is an important question that has never been carefully checked. In this section we discuss how we can use our results to test the gauge invariance of the 3PI and SD formalisms that we use, in spite of the fact that we have done all calculations in Landau gauge. 

The reason that Landau gauge is easiest to use is that the two point functions depend only on the transverse vertex dressing functions ($\tau_i$), and the integral equations for the $\tau_i$ are completely decoupled from the longitudinal components ($\lambda_i$). This means that we can set $\lambda_i$ to zero and still obtain solutions for the two point functions and transverse vertex components. Furthermore, we could avoid calculating the $\lambda_i$ altogether by using the Ward identity to obtain them.  To do this we substitute $S^{-1}$ and $\Gamma_\mu$ (eqs.~(\ref{sigma-eq}, \ref{Pieqn})) into the Ward identity which is
\be
-i q_\mu \Gamma_\mu(p,k) = S^{-1}(k) - S^{-1}(p)\,. \label{wardIdentity}
\ee
The transverse vertex components are immediately removed because of the factor $q_\mu$ which acts as a longitudinal projector. We construct three scalar equations by multiplying by $(1,\slashed{p},\slashed{k})$, tracing, and solving these equations for the longitudinal components. This gives 
\begin{subequations}
\begin{equation}
 \lambda_1 = \frac{ A(k)+ A(p)}{2q^2} \label{wi-components-a}
\end{equation}
\begin{equation}
\lambda_{2} = \frac{B(k) - B(p)}{(k^2-p^2)q^2} \label{wi-components-b} 
\end{equation}
\begin{equation}
\lambda_{3} = \frac{A(k)-A(p)}{2(k^2-p^2)q^2} \,.\label{wi-components-c}
\end{equation}
\end{subequations}
We can check the gauge invariance of the 3PI and SD methods by calculating the longitudinal dressing functions from their self-consistent integral equations and comparing with the expressions in (\ref{wi-components-a} - \ref{wi-components-c}). 
These results are shown in section \ref{sec-gaugeinv}. 

It is also interesting to compare the non-perturbative vertex with the BC vertex ansatz which is used extensively in the literature because it is so much easier to work with. 
The BC ansatz is constructed using only one constraint, which is that it must satisfy the Ward identity (\ref{wardIdentity}). We also know that if the vertex had transverse components, they would disappear when the vertex is contracted with $q_\mu$. These two things together appear to indicate that the BC vertex is longitudinal but this is not true. 
The BC vertex is 
\bea
\Gamma_{\rm BC}^\mu = \gamma^\mu+\frac{1}{2}(A(p)+A(k))\gamma^\mu
+\left[\frac{1}{2}(\slashed{p}+\slashed{k})(A(p)-A(k))+i(B(p)-B(k))\right]\frac{p^\mu+k^\mu}{p^2-k^2}\,
\nn \\\label{bc}
\eea
and it satisfies the Ward identity (\ref{wardIdentity}) which can be seen directly by calculating $q^\mu\Gamma^\mu_{\rm BC}$. 
It is however not the only possible vertex ansatz that satisfies the WI. For example, our non-perturbative vertex with the transverse components set to zero and the longitudinal dressing functions defined as in (\ref{wi-components-a} - \ref{wi-components-c}) also satisfies the WI. 

To understand the relationship between the BC vertex ansatz and the 3PI or SD vertex we first rewrite the BC vertex in the form
\bea
\Gamma_{\rm BC}^\mu &=& \gamma^\mu + \sigma_{1} S^\mu_{1}+ \sigma_{2} S^\mu_{2}+ i \sigma_{3} S^\mu_{3} \label{BCvert} 
\eea
with
\bea
S^\mu_{1} &=&\gamma^\mu \nn \\
S^\mu_{2} &=& \left(\slashed{p}+\slashed{k} \right)\left( p^\mu+k^\mu\right)\nn  \\
S^\mu_{3} &=& \left(p^\mu+k^\mu\right) 
\eea
and
\bea
\sigma_{1} &=& \frac{1}{2}\left(A(k) + A(p) \right) \nn \\
\sigma_{2} &=& \frac{\left(A(k)-A(p)\right)}{2\left(k^2-p^2\right)} \nn \\
\sigma_{3} &=& \frac{B(k)-B(p)}{k^2-p^2} \,.\label{BClam3} 
\eea
Solving the set of equations
\bea
{\rm Tr }\left[L^\mu_{i} \left(\Gamma_\mu - \Gamma^{BC}_{\mu}\right)\right] &=& 0 \nn \\
{\rm Tr }\left[T^\mu_{i} \left(\Gamma_\mu - \Gamma^{BC}_{\mu}\right)\right] &=& 0 \label{BCsoleqn}
\eea
gives
\bea
\frac{\tau_1}{2} &=& \lambda_2 = \frac{\sigma_3}{q^2} \nn \\
\frac{\tau_2}{2} &=& \lambda_3 = \frac{\sigma_2}{q^2} \nn \\
\tau_3 &=& \lambda_1 = \frac{\sigma_1}{q^2} \nn \\[1mm]
\tau_4 &=& \tau_5= \tau_6= \tau_7= \tau_8= \lambda_4= 0 \,.\label{BClam1asOurs}
\eea
It is easy to see that substituting (\ref{BClam3}) into (\ref{BClam1asOurs}) reproduces (\ref{wi-components-a} - \ref{wi-components-c}) which shows that the BC ansatz correctly gives the longitudinal components that satisfy the Ward identity, as it was designed to. One also sees however that the first three transverse components are required to have a specific form, and the last five are set to zero. To test the accuracy of the BC vertex one can check how well eq.~(\ref{BClam1asOurs}) is satisfied by the 3PI and SD vertices. These results are shown in sec.~\ref{sec-BCansatz}.

We also comment that an even simpler ansatz is often used in which only the first term in eq.~(\ref{BCvert}) is kept 
\bea
\Gamma_{{\rm BC}_1}^\mu = \gamma^\mu+\frac{1}{2}(A(p)+A(k))\gamma^\mu \,.
\label{BC1}
\eea
This vertex is not gauge invariant. It is known to agree well with the full BC vertex in some cases \cite{Carrington:2016fsh} but in ref.~\cite{Carrington:2022lzi} it was shown that for an anisotropic system the approximation made in eq.~(\ref{BC1}) can significantly change results. The motivation for using (\ref{BC1}) is that the $p\to k$ limit in the self-consistent integral equations for dressing functions must be treated carefully for the terms in the full BC vertex that are not kept in (\ref{BC1}).
This means that it is much easier to find numerical solutions using the approximation in (\ref{BC1}). We show a comparison of the full BC vertex and the truncated version of it in sec.~\ref{sec-BCansatz}. 

\section{Simplification and Numerical Method}
\label{sec-method}

As discussed in sec.~\ref{scalar-sec}, an important aspect of the numerical calculation is the method used to simplify the set of self-consistent equations that we need to solve, before attempting to find a numerical solution. The basic problem is the appearance in denominators of factors of the form $q^2 = k^2 + p^2 - 2 k \cdot p$ and $\Delta = k^2 p^2 - (k \cdot p)^2$, and powers of these expressions. 
The full phase space of the vertex dressing functions includes points where $q^2$ and $\Delta$ are zero.
The dressing functions obtained from the basis in (\ref{our-basis}) are analytically finite at these points, but numerically we cannot allow summations to include terms of the form 0/0. These problem points can be avoided (for example, by using minimum and maximum values of $\theta$ that are slightly shifted from $0$ and $\pi$), but this does not completely solve the problem. Numerically using $\theta\in(\epsilon,\pi-\epsilon)$ still involves division of small numbers by other small numbers, which means that delicate cancellations are needed to obtain a numerically accurate result. 

It is crucial to simplify the set of self-consistent equations we are trying to solve before attempting to find a numerical solution. To understand how these simplifications work consider the simple example of a large sum of terms that contain the three terms 
\bea
\frac{\chi B_{p+r} p^4}{q^4} 
+ \frac{\chi k^2 B_{p+r}p^2}{q^4}
-\frac{2k \chi B_{p+r} p^3 \cos(\theta)}{q^4} 
= \frac{\chi B_{p+r} p^2}{q^2}
\label{cleaner-example}
\eea
where $\chi$ denotes a common prefactor that is any product of other dressing functions and momenta. When these three terms are combined the power of $q^2$ in the denominator is smaller and the number of terms in the integrand is reduced. Numerically, when $q$ is small three very large terms with different signs are replaced by one much smaller term. We implement a procedure to identify sets of terms that can be rewritten as a smaller number of smaller terms (as in eq.~(\ref{cleaner-example})). This is done in Mathematica using the Simplify function. The basic method is to isolate sets of terms with similar prefactors, apply the function Simplify, and see if a grouping is produced that will cancel with a factor in the denominator. It is important to remember that Mathematica uses implicit alphabetic ordering of variables, which means that $p^2$, $k^2$, and $k \cdot p$ will not necessarily be consecutively ordered within some group of variables. To help Mathematica to recognize variables that should be combined, we relabel $p^2\to x_1$ and $k^2\to x_2$ and $k\cdot p \to x_3$ so that the factors we are interested in are alphabetically adjacent.

The main steps of the algorithm we use to simplify integrands are  as follows.

\begin{enumerate}
\item Collect terms with the same powers of all factors except $p^2$, $k^2$ and $k\cdot p$.
\item Apply Mathematica's Simplify function and identify combinations of $p^2$, $k^2$, and $k \cdot p$ that  give either $q^2$ or $\Delta$.  
\item Repeat steps $1$ and $2$ until the number of terms of the integrand is not reduced. Note that it is necessary to repeat the process because a simplified term has a different power of $q^2$ or $\Delta$ than the terms it was made from, like in (\ref{cleaner-example}), which means that in step $1$ the new term will be collected with a different set and might again find partners it can be combined with.

\item Perform the collection in step $1$ then select terms where the denominator has at least one factor of $q^2$ and the numerator has at least one factor of $k \cdot p$. Replace $k \cdot p$ with $\frac{1}{2}\left(k^2+p^2-q^2\right)$ and Expand. If the length of the integrand does not increase, keep the new expression, if it does, revert to the original. The new expression (if kept) will have at worst the same number of terms, one of which has a smaller power of $1/q^2$.

\item Repeat step $4$ until the length of the integrand is no longer reduced.
\item Perform steps $4$ and $5$ but with $\Delta$ in the denominator instead of $q^2$ and 
$(k \cdot p)^2$ in the numerator instead of $k \cdot p$, 
and replace $\left(k \cdot p\right)^2$ with $k^2p^2-\Delta$. 

\item Go back to step $1$ and repeat the entire process until the length of the integrand does not change.
\end{enumerate}

For example, for the integral equation that gives $\tau_2$, the original expression has $ 629,221$ terms and each of them have at least one power of $q^2$ or $\Delta$ in the denominator. The number with each of these factors is shown in table \ref{cleaner-tau2} in the column labelled `before.' 
After applying the algorithm described above the total number of terms is reduced to $106,034$ and, much more importantly, the number of terms with potentially very small denominators is reduced by an even greater factor. This is shown in the column of the table labelled `after.' To quantify the advantage obtained by the simplification procedure we define a weighting function as 
\bea
w = \sum_i \left[-\text{EX}(\text{term}_i,q^2)-2\,\text{EX}(\text{term}_i,\Delta)\right]
\eea
where the sum is over all terms in a given vertex function and the notation EX(term$_i$,$X$\_) indicates the exponent of $X$ for term$_i$. For example, a term with denominator $q^2\Delta$ contributes 3 to $w$, a term with denominator $q^2\Delta^2$ contributes 5, etc. 
The weight assigned to factors of $\Delta$ is larger than the weight assigned to $q^2$ because $\Delta$ goes to zero for $k\parallel p$ whereas $q^2=0$ requires $k^\mu=p^\mu$. The ratio of the values of $w$ before and after simplification is between $1.435$ (for $\tau_8$) and $9.002$ (for $\tau_2$). 
Table I shows the counts for each type of denominator and the values of $w$ for $\tau_2$ integrand and illustrates how the simplification process  works to remove terms with denominators that can be small. 
\begin{table}[H]
	\begin{center}
		\renewcommand{\arraystretch}{1.5}
		\begin{tabular}{|c|c|c|}
			\hline
			~~~factor~~~ & ~~~~~~~~~~before~~~~~~~~~~  &  ~~~~~~~~~~after~~~~~~~~~~\\
			\hline
			$1$	            & $0 $ & $5,645 $ \\
			\hline
			$q^2$	        & $0 $ & $193 $ \\
			\hline
			$\Delta$	    & $0 $ & $43,705 $ \\
			\hline
			$q^2 \Delta$	& $0 $ & $2,631 $  \\
			\hline
			$\Delta^2$	    & $333,530 $ & $52,352 $  \\
			\hline
			$q^2\Delta^2$	& $295,691 $ & $1,508 $ \\
			\hline
			$w$	& $2,812,575 $ & $312,444 $ \\
			\hline
			$w$ ratio	& \multicolumn{2}{|c|}{$\sim 9.002$} \\
			\hline
			\end{tabular}
		\caption{Efficiency of the simplification process for the dressing function $\tau_2$.}
		\label{cleaner-tau2}
	\end{center}
\end{table}


After all integrands have been simplified as much as possible, we can evaluate the self-consistent equations for the dressing functions numerically. 
Numerically momentum integrals have to be calculated with a finite cutoff $\Lambda$. We scale all variables and functions by the appropriate value of $\Lambda$ based on their mass dimension. The only remaining dimensionful scale is the coupling constant (which is not dimensionless in three dimensional QED).
In all calculations we set the fermion mass to 1 in units of the cutoff and add a regulator $10^{-5}$ to the photon dressing function to prevent a numerical instability at very small momentum. The dependence on the fermion mass is an interesting question but will be left for a future publication. 
We also use symmetry to shorten the calculation. All of the vertex dressing functions are even under an interchange of the external momenta $p$ and $k$, other than $\tau_6$, which is odd. This symmetry allows us to calculate only the region $p \leq k$ and then fill in the function values where $p > k$. 
We calculate all integrals using Gauss-Legendre quadrature.

Another issue involved in the numerical solution of a set of self-consistent integral equations is the interpolation method used to find values of the dressing functions inside the integrals from the calculated arrays, at each successive iteration of the convergence procedure. 
We used linear interpolation. We tested the use of Pad\'e-approximants and found that although they were more accurate than linear interpolation when using the same number of grid points, as expected, it was faster to achieve the same accuracy by using linear interpolation with a larger number of grid points. 
Our grid consists of $33$ values for each external variable (for vertex dressing functions these are $\{p,k,\theta\}$ in our notation) and 16 values for integration variables ($r$ and two angular variables for vertex dressing functions). We tested these values by checking that all results are stable when the number of grid points is increased (see sec.~\ref{sec-res-prelim}).

The convergence procedure we use to solve the set of self-consistent integral equations requires that we start with an initial guess for all dressing functions, solve the self-consistent equations using this initialization, substitute the solutions and recalculate, repeating until the solution for each dressing function does not change. Numerically this is defined as the iteration for which the relative difference with the previous iteration is less than $10^{-4}$. 
We define the relative difference of two arrays $f$ and $g$ by the root-mean-square
\bea
&& \bar{r}(f,g) =\sum_i^N \frac{\sqrt{{\rm rel}(f,g)_i^2}}{N} \nn \\
&& {\rm rel}(f,g)_i = \left| \frac{2\left(f(x_i)- g(x_i)\right)}{f(x_i)+g(x_i)} \right|\,\label{rel}
\eea
where $x_i$ represents variables $(p,k,\theta)$ and $N$ is the total number of grid points in the $p$, $k$ and $\theta$ dimensions. Points for which the function is less than $10^{-6}$ are not compared in order to avoid large relative differences produced by checking the accuracy of very small points.  

After each iteration we update each dressing function using a damping factor as follows. Starting from the results for iteration $(i-1)$, denoted $F_{i-1}(k,p)$ for any dressing function $F$, we calculate the results $F_{i}(k,p)$ of iteration $i$ and then redefine
\be
F_i(k,p) = 0.8 F_{i}(k,p) + 0.2 F_{i-1}(k,p) \label{damping}
\ee
so the updated functions are 80\% the new result and 20\% the old one. 
In many cases the use of this procedure means more iterations will be required to find a converged solution. 
However, damping prevents the occurrence of oscillatory behaviour in cases where a dressing function, typically with small values at all points in its phase space, effectively oscillates between sets of values that bracket the true solution. We have found that the fastest method overall is to implement this damping factor on all dressing functions. 

The convergence of the set of self-consistent equations is obviously much faster if the initialization is close to the solution. To maximize efficiency we use the following procedure. We start with a small coupling ($\alpha=0.64$) and initialize with the one loop forms of all dressing functions. The full set of self-consistent equations converged in 21 iterations. We repeated the calculation with $\alpha=1$ starting from the converged results obtained with $\alpha=0.64$. In the same way we solved the equations with $\alpha=2$ starting from the $\alpha=1$ solutions, and finally $\alpha=5$ starting from the $\alpha=2$ solutions. The $\alpha=1$ results were obtained in 14 iterations and $\alpha=5$ in  35 iterations. 

We use Message Passing Interface (MPI) to parallelize the program. Most of the time cost of our program is due to the nested integral loops which do not need to share data across external momenta other than the completed integral values themselves. This means our program scaled very favourably with a high number of parallel processes. Typically one iteration takes 1.7 hours using 561 cores. 

\section{Results}
\label{sec-results}
\subsection{Preliminary}
\label{sec-res-prelim}

One of the main goals of this work is to compare the extent to which the 3PI and SD vertex functions satisfy the Ward identity. This gives an indication of the extent to which each method can be trusted. We will also test how well the BC vertex ansatz agrees with the non-perturbative vertex functions. This is important information because the BC vertex is used extensively in the literature without any way to test its accuracy. Finally we will evaluate which of the vertex dressing functions are the largest and therefore the most important. This information could provide a useful way to simplify future calculations. If one of the dressing functions is consistently very small, then setting that dressing function to zero will produce a reliable approximation to the full vertex at reduced computational cost. Alternatively one could drop the smallest dressing functions from the full set of equations, converge the remaining dressing functions, and then reinstate the small functions at the end and reconverge. This type of piecewise convergence can have significant numerical advantages in situations where very high accuracy is needed.

In sec.~\ref{sec-method} we explained how we converged the 3PI or SD integral equations by iterating their solutions until the relative difference with the previous iteration was smaller than $10^{-4}$ (see eq.~(\ref{rel})). In this section we use a slightly different method to compare two arrays. 
We want to compare arrays that would be very close to each other if the Ward identity were satisfied or if the BC ansatz was an accurate representation of the full 3PI or SD vertex.  The two arrays that are being compared can in fact look very different from each other, and this means that we need to take their absolute values before calculating the relative difference. If this is not done, artificially large results are obtained when there are a small number of points for which the two arrays that are being compared have opposite sign. The number of points for which this happens is usually very small. 

The dimensions of the bi-spinors/tensors in our basis (\ref{our-basis}) are not all the same, which means that the dimensions of the dressing functions are not the same. To compare the relative importance of different vertex components we define dimensionless functions (denoted with hats) as
\bea
&& \hat \tau_1 = \frac{1}{2} \sqrt{q^2}\left(k^2-p^2\right) \tau_1 \nn \\
&& \hat \tau_2 = \frac{1}{2} q^2 \left(k^2-p^2\right) \tau_2 \nn \\
&& \hat \tau_3 = q^2 \tau_3 \nn \\
&& \hat \tau_4 = (k p)^{3/2} \left(k^2-p^2\right) \tau_4 \nn \\
&& \hat \tau_5 = \sqrt{k p} \tau_5 \nn \\
&& \hat \tau_6 = \sqrt{k p q^2} \tau_6 \nn \\
&& \hat \tau_7 = \sqrt{k p}\left(k^2-p^2\right)\tau_7 \nn \\
&& \hat \tau_8 = \sqrt{k p q^2}\tau_8 \nn \\
&& \hat \lambda_1 = q^2 \lambda_1 \nn \\
&& \hat \lambda_2 = \sqrt{q^2} \left(k^2-p^2\right) \lambda_2 \nn \\
&& \hat \lambda_3 = q^2 \left(k^2-p^2\right)\lambda_3\,. \label{hat-defn}
\eea
For the BC dressing functions we use 
\bea
\hat{\sigma}_1 &=&  \sigma_1 \\
\hat{\sigma}_2 &=& \left(k^2-p^2\right) \sigma_2\\
\hat{\sigma}_3 &=& \frac{\left(k^2-p^2\right)}{\sqrt{q^2}} \sigma_3\,.
\eea
The factors for the BC dressing functions are chosen from eq.~(\ref{BClam3})
and the factors for the three longitudinal dressing functions and for $(\tau_1,\tau_2,\tau_3)$ are defined to match those of the corresponding $\sigma_i$ according to eq.~(\ref{BClam1asOurs}). The factors for $(\tau_4, \cdots, \tau_8)$ are chosen from inspection of the one loop results in eq.~(\ref{1loopLam3}). 

The Ward identity should be satisfied identically at one loop. 
As a preliminary test of our program we checked the Ward identity in (\ref{wi-components-a} - \ref{wi-components-c}) using the numerical results obtained from our program initialized with all dressing functions set to zero and stopping after the first iteration (which means working at the one loop level). The results of this numerical check are shown in table \ref{tableOneLoop} and verify that our computational method is sufficiently accurate in this limit.

\begin{table}[H]
		\begin{center}
	\begin{tabular}{|c|c|c|c|}
		\hline
~~~	equation	~~~ & ~~~~~~~~~~(\ref{wi-components-a})~~~~~~~~~~  &  ~~~~~~~~~~(\ref{wi-components-b})~~~~~~~~~~ &  ~~~~~~~~~~(\ref{wi-components-c})~~~~~~~~~~  \\
		\hline
		$\bar{r}$	& $3.17 \times 10^{-5}$ &	$ 9.72 \times 10^{-11}$ & $9.82 \times 10^{-10}$ \\
		\hline
	\end{tabular}
	\caption{Accuracy of the one loop dressing functions}
	\label{tableOneLoop}
	\end{center}
\end{table}

We have checked that all results are stable when the number of grid points is increased. In all calculations except those shown in table \ref{table-converg} we use $33$ grid points for each external variable (for vertex dressing functions these are $\{p,k,\theta\}$ in our notation) and 16 values for integration variables ($r$ and two angular variables denoted $\theta_r$ and $\phi_r$ for vertex dressing functions). We denote $N_k=N_p \equiv N_1$ and $N_r=N_\theta=N_{\theta_r}=N_{\phi_r} \equiv N_2$ and $N_T=N_1\times N_2$ so that our calculation corresponds to $N_T=33\times 16 = 528$. In table \ref{table-converg} we show the values of the fermion two point dressing functions at zero momentum (numerically $10^{-2.5}$) and the photon dressing function at $p=1$. 
\begin{center}
\begin{table}[H]
\begin{center}
\begin{tabular}{|c|c|c|c|c|}
\hline
~~~~~$N_T$~~~~~& ~~~~~~~~~348~~~~~~~~~& ~~~~~~~~~~434 ~~~~~~~~~~ &~~~~~~~~~~528~~~~~~~~~~&~~~~~~~~~~630~~~~~~~~~~ \\
\hline
\hline
~~~~~$A(0)$ ~~~~~~ & 0.0110514 & 0.0110748&  0.0111117& 0.0110541 \\
\hline
~~~~~$B(0)$ ~~~~~~ & 0.0488779 & 0.0488819 & 0.0488836 & 0.0488865 \\
\hline
~~~~~$\Pi(1)$ ~~~~~~ & 0.00756199 & 0.00754743 & 0.00753515 & 0.00752456 \\
\hline				
\end{tabular}
\end{center}
\caption{Stability of results with changing number of grid points, see text for description.}
\label{table-converg}
\end{table}
\end{center}

\subsection{Numerical results for dressing functions}

\begin{figure}[H]
	\centering\begin{subfigure}{0.45\textwidth}
		\centering
		\includegraphics[width=0.95\textwidth]{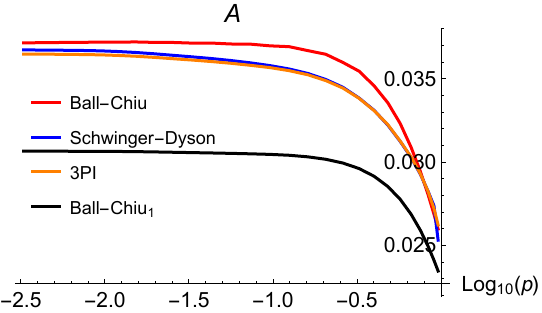}
		\caption{}
	\end{subfigure} 
	\begin{subfigure}{0.45\textwidth}
		\centering
		\includegraphics[width=0.95\textwidth]{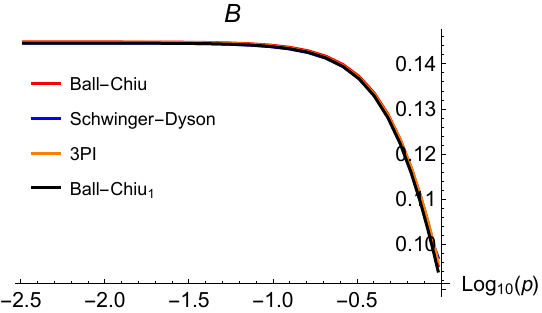}
		\caption{}
	\end{subfigure} 
	\begin{subfigure}{0.45\textwidth}
		\centering
		\includegraphics[width=0.95\textwidth]{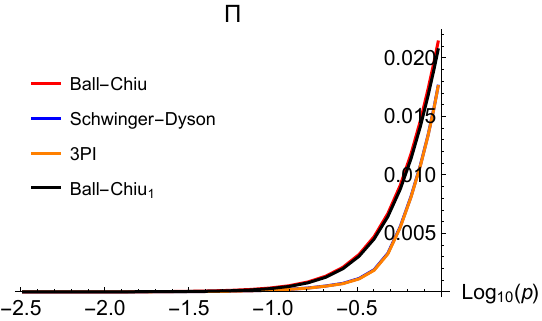}
		\caption{}
	\end{subfigure} 
	\begin{subfigure}{0.45\textwidth}
		\centering
		\includegraphics[width=0.95\textwidth]{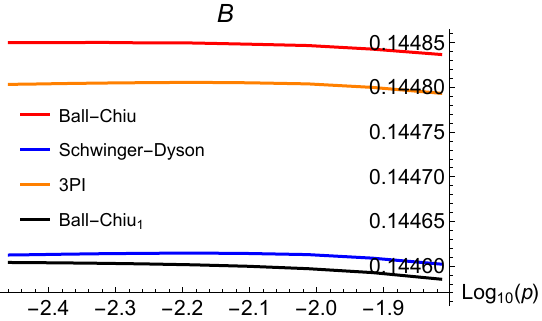}
		\caption{}
	\end{subfigure} 
	\caption{Fermion and photon dressing functions using different vertex methods with $\alpha=2$, and part (d) shows a close up of the small momentum region for $B(p)$. \label{al2-fig}}
\end{figure}

In fig.~\ref{al2-fig} we show the two point functions calculated with coupling $\alpha=2$. 
The calculations are done using the 3PI vertex, the SD vertex, the BC ansatz, and the BC$_1$ 
ansatz in eq.~(\ref{BC1}). The results show that for the fermion dressing function $B$ all four methods agree with each other well. For the function $A$, the 3PI and SD calculations agree well, but BC vertex gives a result that is noticeably larger, and BC$_1$ is even more significantly smaller. For the photon dressing function, the 3PI and SD results agree well with each other, and the BC and BC$_1$ calculations also agree with each other, but the two sets of curves are noticeably different at large momentum. For the function $B$ all calculations agree well. The fourth part of the figure shows a close up of the small momentum region and shows slight differences between the four different calculations. 

\begin{figure}[H]
	\centering\begin{subfigure}{0.45\textwidth}
		\centering
		\includegraphics[width=0.95\textwidth]{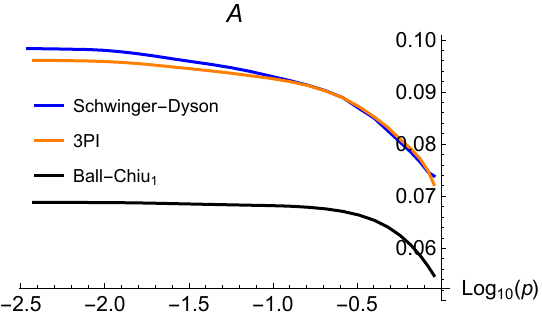}
		\caption{}
	\end{subfigure} 
	\begin{subfigure}{0.45\textwidth}
		\centering
		\includegraphics[width=0.95\textwidth]{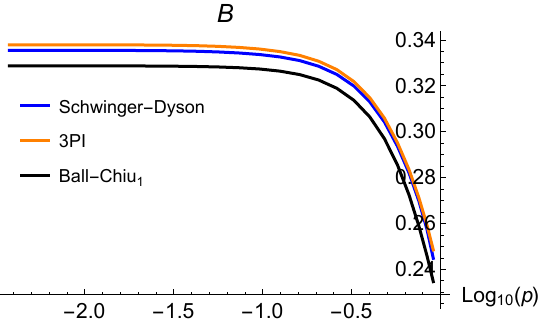}
		\caption{}
	\end{subfigure} 
	\begin{subfigure}{0.45\textwidth}
		\centering
		\includegraphics[width=0.95\textwidth]{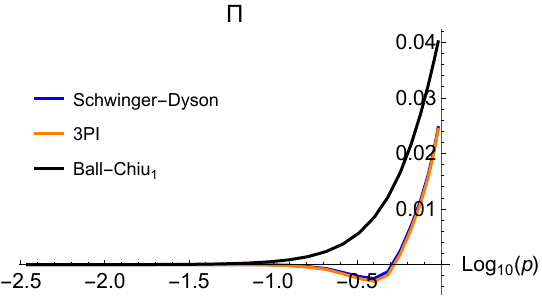}
		\caption{}
	\end{subfigure} 
	\caption{Fermion and photon dressing functions using different vertex methods with $\alpha=5$. \label{al5-fig}}
\end{figure}

One expects that the different methods we are comparing will give more significantly different results when the coupling is larger. In fig.~\ref{al5-fig} we show the two point dressing functions obtained with $\alpha=5$ from the 3PI, SD, and BC$_1$ calculations. The full BC vertex is more numerically unstable than any of the other three calculations, and we were not able to find a solution with $\alpha=5$. The figures show that, as expected, the difference from the three methods is greater at larger coupling, although the 3PI and SD methods agree fairly well, especially at large momentum.

\begin{figure}[H]
			\centering
			\includegraphics[width=0.75\textwidth]{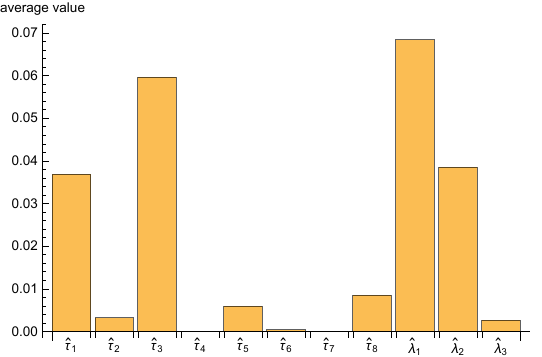}
			\caption{The average value of the vertex dressing functions for $\alpha=5$.}
	\label{avg}
\end{figure}

To give an idea of the size of each vertex dressing function we calculate the average of the absolute value of all points. These results are shown in table \ref{avg} and fig.~\ref{avg}. The largest three,  $\langle \hat{\lambda}_1 \rangle$, $\langle\hat{\tau}_3\rangle$ and $\langle \hat{\lambda}_2 \rangle$, are bigger than the smallest three, $\langle\hat{\tau}_4 \rangle$, $\langle\hat{\tau}_6 \rangle$, and $\langle\hat{\tau}_7 \rangle$, by a factor of about $100$. 

\begin{figure}[H]
	\centering
	\foreach \i in {1,2,3,4,5,6,7,8} {
		\begin{subfigure}{0.35\textwidth}
			\centering
			\includegraphics[width=0.95\textwidth]{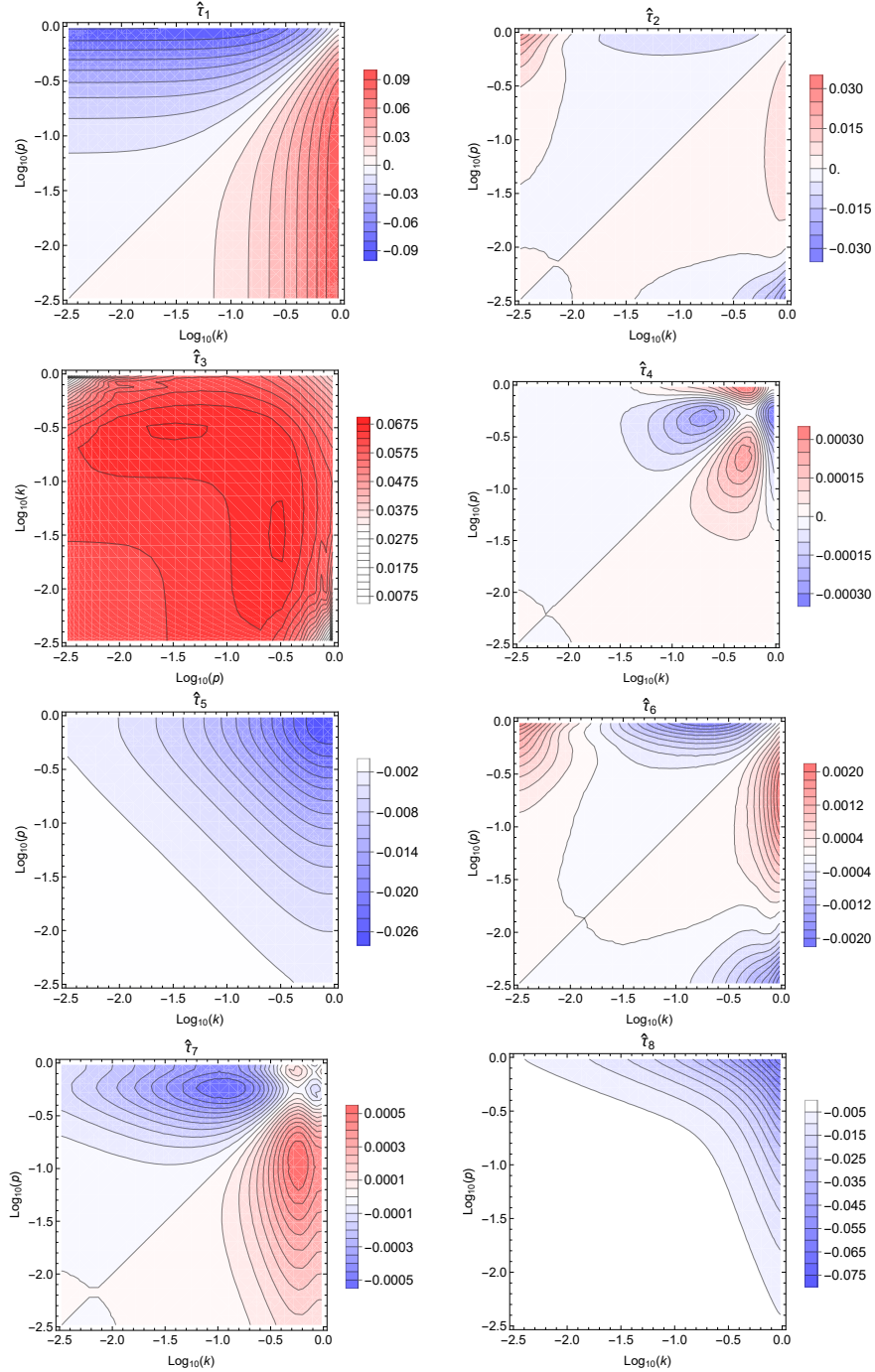}
	\end{subfigure} }
	\caption{Transverse dressing functions with $\alpha=5$ and $\theta = \pi/2$. \label{trans-fig}}
\end{figure}

\begin{figure}[H]
	\centering
	\foreach \i [remember=\myval as \myval (initially 0)] in {10,11,12} {
		\pgfmathsetmacro{\myval}{\myval + 1}
		
		\begin{subfigure}{0.45\textwidth}
			\centering
			\includegraphics[width=0.95\textwidth]{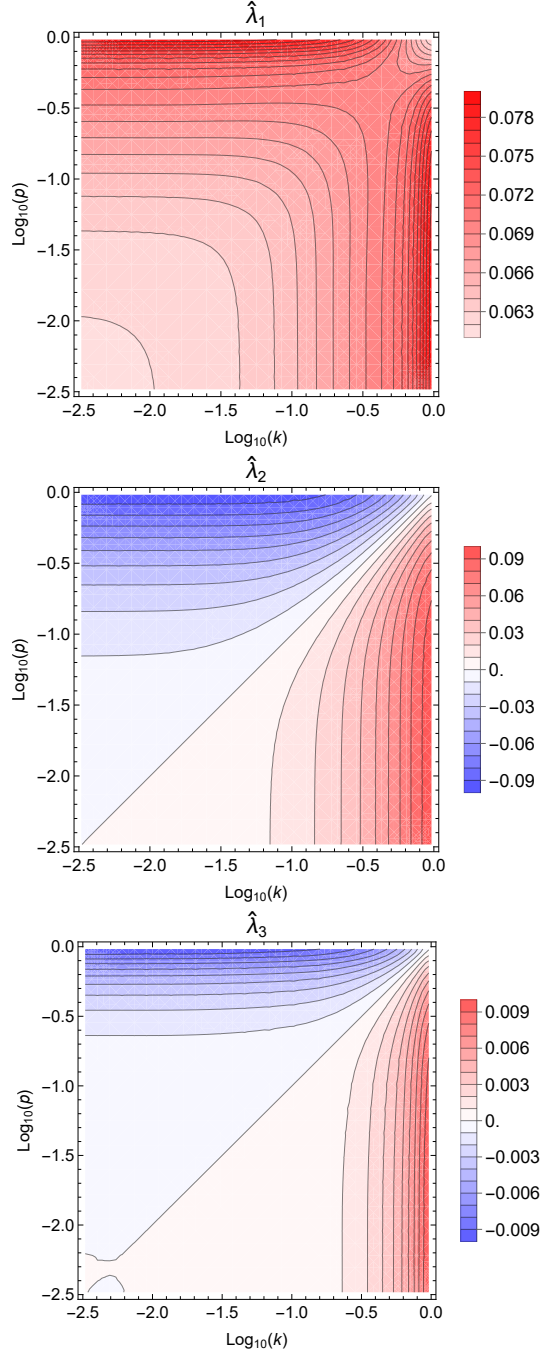}
	\end{subfigure} }
	\caption{Longitudinal dressing functions with $\alpha=5$ and $\theta = \pi/2$. \label{long-fig}}
\end{figure}

In figs.~\ref{trans-fig} and \ref{long-fig} we show the 11 vertex dressing functions with $\alpha=5$ and $\theta=\pi/2$ (numerically $1.5208$). Results depend only weakly on $\theta$. We comment that since the BC ansatz is completely independent of $\theta$ this is a necessary condition for the BC ansatz to able to accurately represent the full vertex. 

\subsection{Gauge invariance}
\label{sec-gaugeinv}
In this section we test the gauge invariance of the 3PI and SD calculations. In table~\ref{LRComp} we show the relative difference of the two sides of eq.~(\ref{wi-components-a} - \ref{wi-components-c}) for different values of coupling. The last column of the table shows a weighted average value for all three dressing functions for each coupling. The weights for each dressing function are calculated from the results in fig.~\ref{avg} and are meant to take into account the fact that a large relative difference for a very small dressing function will probably not have a correspondingly large effect on something physical, like the fermion condensate $B(0)$. The results in table \ref{LRComp} show that the violation of the Ward identity increases with coupling and is slightly worse for the SD calculation. 

\begin{center}
	\begin{table}[H]
		\begin{center}
			\renewcommand{\arraystretch}{1.5}
			\begin{tabular}{|c|c|c|c|c|c|}
				\hline
				~~~~coupling~~~~ & ~~method~~ & ~~~~~~~~$\hat\lambda_1$~~~~~~~~ &~~~~~~~~$\hat\lambda_2$~~~~~~~~ & ~~~~~~~~$\hat\lambda_3$~~~~~~~~ & ~~~~weighted average~~~~ \\
				\hline
				\hline
				$\alpha=0.64$&3PI   &$6.08\times 10^{-2}$  &$0.155$   &$0.710$   &$0.113$ \\
				             &SD    &$6.23\times 10^{-2}$  &$0.157$   &$0.746$   &$0.116$ \\
				\hline
				$\alpha=2.0$ &3PI   &$0.153$               &$0.194$   &$0.873$   &$0.186 $ \\
				             &SD    &$0.163$               &$0.188$   &$0.967$   &$0.190 $ \\
				\hline
				$\alpha=5.0$ & 3PI  &$0.271$               &$0.315$   &$1.00$    &$0.305 $ \\
				             & SD   &$0.311$               &$0.285$   &$1.16$    &$0.322 $ \\
				\hline
				\hline
			\end{tabular}
		\end{center}
		\caption{The relative difference $\bar r$ of the left and right sides of (\ref{wi-components-a} - \ref{wi-components-c}) using different methods and couplings. The last column shows a weighted average (see text for description).}
		\label{LRComp}
	\end{table}
\end{center}

\subsection{Accuracy of the BC ansatz}
\label{sec-BCansatz}

We also check how well the BC ansatz agrees with the 3PI and SD non-perturbative vertices. To do this we test how each component of the non-perturbative vertex agrees with the corresponding result from the BC ansatz, using eqs.~(\ref{BClam3}, \ref{BClam1asOurs}). The results are shown in table \ref{BC-table} and show that the BC ansatz agrees reasonably well with the non-perturbative vertex from either  method at very small coupling. At larger coupling the disagreement grows, and is consistently slightly worse for the SD vertex.  

\begin{figure}[H]
\centering
\includegraphics[width=0.376\textwidth]{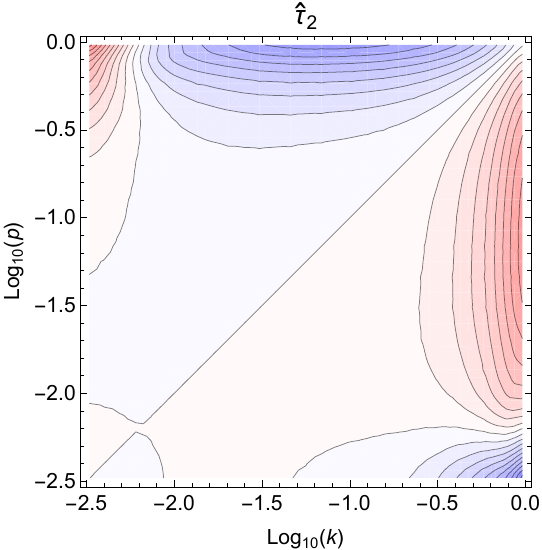}
\includegraphics[width=0.465\textwidth]{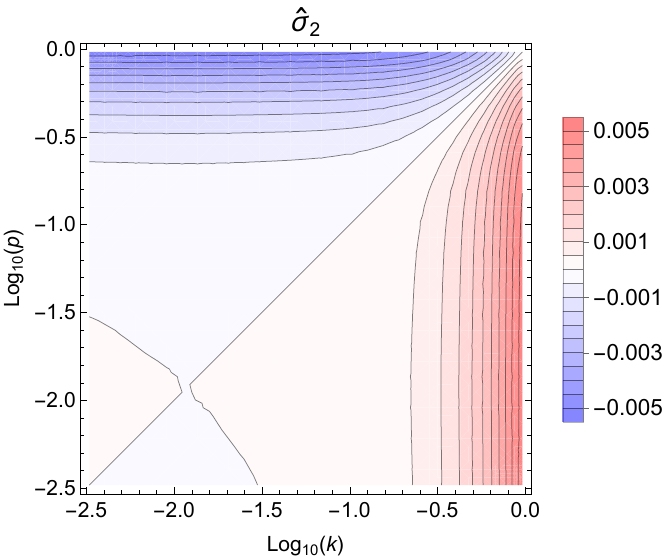}
	\caption{The dressing functions $\hat\sigma_2$ and $\hat\tau_2$ with $\alpha=2$ and $\theta = \pi/2$. \label{tau2sig2-fig}}
\end{figure}

\begin{center}
	\begin{table}[H]
		\begin{center}
			\renewcommand{\arraystretch}{1.5}
			\begin{tabular}{|c|c|c|c|c|}
				\hline
				coupling & \multicolumn{2}{|c|}{$\alpha=0.64$}& \multicolumn{2}{|c|}{$\alpha=2$}\\
				\hline
				~~~~~method~~~~~ & ~~~~~~~~~3PI~~~~~~~~~ & ~~~~~~~~~SD~~~~~~~~~ & ~~~~~~~~~3PI~~~~~~~~~ & ~~~~~~~~~SD~~~~~~~~~ \\
				\hline
				$\bar{r}(\hat{\sigma}_1,\hat{\lambda}_1)$ &$7.46\times 10^{-2}$ &$7.55\times 10^{-2}$  &$0.177$ &$0.184$              \\
				\hline
				$\bar{r}(\hat{\sigma}_2,\hat{\lambda}_3)$ &$0.368$              &$0.404$               &$0.565$ &$0.643$              \\
				\hline
				$\bar{r}(\hat{\sigma}_3,\hat{\lambda}_2)$ &$7.15\times 10^{-2}$ &$8.63\times 10^{-2}$  &$0.155$ &$0.178$              \\
				\hline
				$\bar{r}(\hat{\sigma}_1,\hat{\tau_3})$    &$0.240$              &$0.244$               &$0.332$ &$0.347$              \\
				\hline
				$\bar{r}(\hat{\sigma}_2,\hat{\tau}_2)$    &$0.589$              &$0.638$               &$0.673$ &$0.735$              \\
				\hline
				$\bar{r}(\hat{\sigma}_3,\hat{\tau}_1)$ &$8.78\times 10^{-2}$ &$8.76\times 10^{-2}$  &$0.176$ &$0.171$              \\
				\hline
				\hline
				weighted			&$0.288$              &$0.308 $              &$0.449 $&$0.497 $\\
				average of $\bar{r}$ & & & & \\                   
				\hline
				\hline				
				$\langle \hat{\tau}_4 \rangle$    & $2.88\times 10^{-3}$ & $2.88\times 10^{-3} $ & $7.55\times 10^{-3} $ & $7.61\times 10^{-3} $ \\
				\hline
				$\langle  \hat{\tau}_5 \rangle$    & $2.13\times 10^{-5}$ & $2.05\times 10^{-5} $ & $6.61\times 10^{-5} $ & $6.13\times 10^{-5} $ \\
				\hline
				$\langle  \hat{\tau}_6 \rangle$    & $7.38\times 10^{-4}$ & $7.35\times 10^{-4} $ & $2.18\times 10^{-3} $ & $2.14\times 10^{-3} $ \\
				\hline
				$\langle \hat{\tau}_7 \rangle$    & $1.48\times 10^{-6}$ & $8.09\times 10^{-6} $ & $3.93\times 10^{-6} $ & $5.62\times 10^{-5} $ \\
				\hline
				$\langle \hat{\tau}_8 \rangle$    & $2.29\times 10^{-4}$ & $2.32\times 10^{-4} $ & $6.06\times 10^{-4} $ & $6.30\times 10^{-4} $ \\
				\hline
			\end{tabular}
		\end{center}		
		\caption{Accuracy of eqs.~(\ref{BClam3}, \ref{BClam1asOurs})  using different methods and couplings.}
		\label{BC-table}
	\end{table}
\end{center}
The agreement of $\hat\sigma_2$ and $\hat\tau_2$ is the worst. In fig.~\ref{tau2sig2-fig} we show a contour plot of $\hat\sigma_2$ and $\hat\tau_2$ at $\theta=\pi/2$, with $\alpha=2$. The figure shows that there are significant structural differences between the two vertices. 

\section{Conclusions}
\label{sec-conclusions}

We have studied different methods to calculate non-perturbative vertex dressing functions in strongly coupled QED in three Euclidean dimensions. 
We have solved the equations of motion obtained from the three loop 3PI effective action. 
We have also solved the coupled set of QED SD equations truncated with a bare four point vertex. 
We have compared these results with two commonly used vertex ans\"atze called the Ball-Chiu vertex, and a simpler approximation to this vertex denoted BC$_1$. We have worked exclusively in Landau gauge and assessed the gauge dependence of the formulations we study by checking the extent to which the Ward identity is satisfied. 
We have found that at the largest coupling we used ($\alpha=5$) the Ward identity is violated by approximately 30\% using either the 3PI method or the SD approach.  It is important to remember that the effect of this gauge dependence on a physical quantity, like the fermion condensate ($B(0)$ in our notation), can only be tested by performing calculations in a different gauge. It is encouraging to note that fig.~\ref{al2-fig} shows that the results for the two point functions are in fairly good agreement for all methods, and fig.~\ref{al2-fig}d shows that $B(0)$ from the BC vertex, which explicitly satisfies the Ward identity, and the 3PI equations of motion, are very close together. This suggest that the result of a calculation of a physical quantity like the critical coupling for chiral symmetry breaking could be largely gauge independent. 
A direct check by comparison of the Landau and Feynman gauge results is in progress and will be presented in a future publication. 

\appendix

\section{One loop structure}
\label{ap-1loop}

To do the calculations in this paper it is important to understand the structure of the non-perturbative equations in the perturbative limit. The structure of the basis (\ref{our-basis}) was originally understood by studying the perturbative limit and identifying the combinations of the tensors in (\ref{naif}) that give divergence free dressing functions \cite{pennington-2}. To check our equations we have reproduced these results. 
In our notation (see equations~(\ref{Geqn0},\ref{S-nonpert},\ref{Gam})) all 15 dressing functions go to zero in the perturbative limit, which means we can obtain the perturbative equations from our non-perturbative expressions simply by setting all dressing functions to zero. Alternatively one can calculate the diagrams in figs.~\ref{fermionSelfEnergy},~\ref{photonSelfEnergy} and \ref{vertexCorrection} using bare propagators and vertex functions. The agreement of these results gives a simple check of the procedure we use to generate the non-perturbative equations. The integrals that are produced can be calculated analytically using dimensional regularization. The results are given in arbitrary dimension in Minkowski space in \cite{davy} and a numerical analysis of the one-loop quark-gluon vertex can be found in \cite{Bermudez:2017bpx}. We give below some results in three dimensions and in Euclidean space. We use the notation
\be
{\rm I}\left[ \nu_1,\nu_2,\nu_3 \right] =\int \frac{d^3 r}{(2\pi)^3} \frac{1}{\left((p+r)^2+m^2\right)^{\nu_1} \left((k+r)^2+m^2\right)^{\nu_2} (r^2)^{\nu_3}}.
\ee
The one loop results for the dressing functions in the fermion propagator and the longitudinal vertex functions are
\bea
\delta A(p) &=& \frac{(p^2+m^2)^2{\rm I}[1,0,2]-{\rm I}[1,0,0]}{2 p^2} \nn \label{1loopA} \\
\delta B(p) &=& 2 m {\rm I}[1,0,1] \nn \\
\lambda_1(k,p) &=& \frac{\left(k^2+m^2\right)^2{\rm I}[0,1,2] -{\rm I}[0,1,0]}{4 k^2 q^2}+\frac{\left(p^2+m^2\right)^2{\rm I}[1,0,2]
	-{\rm I}[1,0,0]}{4 p^2 q^2}\nn \\
\lambda_2(k,p) &=& \frac{2 m \left({\rm I}[0,1,1] - {\rm I}[1,0,1]\right)}{q^2 (k^2-p^2)}\nn \\
\lambda_3(k,p) &=& \frac{\left(k^2+m^2\right)^2{\rm I}[0,1,2] -{\rm I}[0,1,0]}{4 k^2 q^2 (k^2-p^2)}-\frac{ \left(p^2+m^2\right)^2{\rm I}[1,0,2]-{\rm I}[1,0,0]}{4
	p^2 q^2 (k^2-p^2)} \label{1loopLam3}
\eea

It is easy to verify that eqs.~(\ref{1loopLam3}) satisfy the Ward identity in the form (\ref{wi-components-a} - \ref{wi-components-c}).

\section*{Acknowledgements}

This work was partially supported by the Natural Sciences and Engineering Research Council of Canada under grants SAPIN-2023-00023 and SAPIN-2020-00054. 



\begin{thebibliography}{99}


\bibitem{BallChiu1}
J.~S.~Ball and T.~W.~Chiu,
Phys. Rev. D \textbf{22}, 2542 (1980).


\bibitem{Roberts:1994dr}
C.~D.~Roberts and A.~G.~Williams,
Prog. Part. Nucl. Phys. \textbf{33}, 477 (1994).


\bibitem{Binosi:2009qm}
D.~Binosi and J.~Papavassiliou,
Phys. Rept. \textbf{479}, 1 (2009).


\bibitem{Calzetta:2004sh}
E.~A.~Calzetta,
Int. J. Theor. Phys. \textbf{43}, 767 (2004).


\bibitem{Reinosa:2009tc}
U.~Reinosa and J.~Serreau,
Annals Phys. \textbf{325}, 969 (2010).


\bibitem{Arrizabalaga:2002hn}
A.~Arrizabalaga and J.~Smit,
Phys. Rev. D \textbf{66}, 065014 (2002).


\bibitem{Carrington:2003ut}
M.~E.~Carrington, G.~Kunstatter and H.~Zaraket,
Eur. Phys. J. C \textbf{42}, 253 (2005).


\bibitem{Brown:2016vaj}
M.~J.~Brown and I.~B.~Whittingham,
Phys. Rev. D \textbf{95}, 025018 (2017).


\bibitem{Huber:2018ned}
M.~Q.~Huber,
Phys. Rept. \textbf{879}, 1 (2020).



\bibitem{pennington0}
D.~C.~Curtis and M.~R.~Pennington,
Phys. Rev. D \textbf{42}, 4165 (1990).


\bibitem{Curtis:1993py}
D.~C.~Curtis and M.~R.~Pennington,
Phys. Rev. D \textbf{48}, 4933 (1993).


\bibitem{Kizilersu:2009kg}
A.~Kizilersu and M.~R.~Pennington,
Phys. Rev. D \textbf{79}, 125020 (2009).


\bibitem{Goecke:2008zh}
T.~Goecke, C.~S.~Fischer and R.~Williams,
Phys. Rev. B \textbf{79}, 064513 (2009).


\bibitem{Kizilersu:2014ela}
A.~K{\i}z{\i}lers{\"u}, T.~Sizer, M.~R.~Pennington, A.~G.~Williams and R.~Williams,
Phys. Rev. D \textbf{91}, 065015 (2015).


\bibitem{Eichmann:2021zuv}
G.~Eichmann, J.~M.~Pawlowski and J.~M.~Silva,
Phys. Rev. D \textbf{104}, 114016 (2021).


\bibitem{Albino:2022efn}
L.~Albino, A.~Bashir, A.~J.~Mizher and A.~Raya,
Phys. Rev. D \textbf{106}, 096007 (2022).


\bibitem{Carrington:2004sn}
M.~E.~Carrington,
Eur. Phys. J. C \textbf{35}, 383 (2004).


\bibitem{Berges:2004pu}
J.~Berges,
Phys. Rev. D \textbf{70}, 105010 (2004).


\bibitem{pennington-2}
A.~Kizilersu, M.~Reenders and M.~R.~Pennington,
Phys. Rev. D \textbf{52}, 1242 (1995).


\bibitem{Eichmann:2026ttr}
G.~Eichmann,
[arXiv:2603.00804 [hep-ph]].


\bibitem{davy}
A.~I.~Davydychev, P.~Osland and L.~Saks,
Phys. Rev. D \textbf{63}, 014022 (2001).


\bibitem{Carrington:2016fsh}
M.~E.~Carrington, C.~S.~Fischer, L.~von Smekal and M.~H.~Thoma,
Phys. Rev. B \textbf{94}, 125102 (2016). 


\bibitem{Carrington:2022lzi}
M.~E.~Carrington, A.~R.~Frey and B.~A.~Meggison,
Phys. Rev. D \textbf{107}, 056012 (2023).


\bibitem{Williams:2015cvx}
R.~Williams, C.~S.~Fischer and W.~Heupel,
Phys. Rev. D \textbf{93}, 034026 (2016).


\bibitem{anapole}
Zel’Dovich, Ia B,
Sov. Phys. JETP \textbf{6}(6), 1184--1186 (1958).


\bibitem{Carrington:2007ea}
M.~E.~Carrington and E.~Kovalchuk,
Phys. Rev. D \textbf{77}, 025015 (2008).


\bibitem{form}
J.~Kuipers, T.~Ueda, J.~A.~M.~Vermaseren and J.~Vollinga,
Comput. Phys. Commun. \textbf{184}, 1453-1467 (2013)


\bibitem{Bermudez:2017bpx}
R.~Bermudez, L.~Albino, L.~X.~Guti{\'e}rrez-Guerrero, M.~E.~Tejeda-Yeomans and A.~Bashir,
Phys. Rev. D \textbf{95}, 034041 (2017).

\end{thebibliography}
\end{document}